\documentclass{article}
\usepackage[utf8]{inputenc}
\usepackage{authblk}
\usepackage{hyperref}
\usepackage{setspace}
\usepackage[margin=1.25in]{geometry}
\usepackage{graphicx}
\graphicspath{ {./figures/} }
\usepackage{subcaption}
\usepackage{float}
\usepackage{amsmath}
\usepackage{graphicx}
\usepackage{hyperref}
\usepackage{adjustbox} 
\usepackage{threeparttable} 

\usepackage[style=vancouver, 
citestyle=numeric-comp, 
sorting=none, 
doi=true, 
url=false, 
isbn=false]{biblatex}
\title{\textbf{Real-Time Brain Tumor Detection in Intraoperative Ultrasound Using YOLO11: From Model Training to Deployment in the Operating Room}}

\author[1,2*$\dag$]{Santiago Cepeda}
\author[1,2$\dag$]{Olga Esteban-Sinovas}
\author[3]{Roberto Romero}
\author[4]{Vikas Singh}
\author[4]{Prakash Shett}
\author[4]{Aliasgar Moiyadi}
\author[5,6]{Ilyess Zemmoura}
\author[7]{Giuseppe Roberto Giammalva}
\author[8,9]{Massimiliano Del Bene}
\author[8]{Arianna Barbotti}
\author[8,10,11]{Francesco DiMeco}
\author[12]{Timothy R. West }
\author[12]{Brian V. Nahed}
\author[1,2]{Ignacio Arrese}
\author[2,13,14]{Roberto Hornero}
\author[1,2]{Rosario Sarabia}

\affil[1]{Department of Neurosurgery, Rio Hortega University Hospital, 47014 Valladolid, Spain}
\affil[2]{Specialized Group in Biomedical Imaging and Computational Analysis (GEIBAC), Instituto de Investigacion Biosanitaria de Valladolid (IBioVALL), 47014, Valladolid, Spain}
\affil[3]{Biomedical Engineering Group, University of Valladolid, 47011 Valladolid, Spain}
\affil[4]{Department of Neurosurgery, Tata Memorial Centre, Homi Bhabha National Institute, Mumbai 400012, Maharashtra, India}
\affil[5]{UMR 1253, iBrain, Université de Tours, Inserm, 37000 Tours, France}
\affil[6]{Department of Neurosurgery, CHRU de Tours, 37000 Tours, France}
\affil[7]{Department of Neurosrugery, ARNAS Civico Di Cristina Benfratelli Hospital, 90127 Palermo, Italy}
\affil[8]{Department of Neurosurgery, Fondazione IRCCS Istituto Neurologico Carlo Besta, Via Celoria 11, 20133 Milan, Italy}
\affil[9]{Department of Pharmacological and Biomolecular Sciences, University of Milan, 20122 Milan, Italy}
\affil[10]{Department of Oncology and Hematology-Oncology, Università Degli Studi di Milano, 20122 Milan, Italy}
\affil[11]{Department of Neurological Surgery, Johns Hopkins Medical School, Baltimore, MD 21205, USA}
\affil[12]{Department of Neurosurgery, Massachusetts General Hospital, Mass General Brigham, Harvard Medical School, Boston, MA 02114, USA}
\affil[13]{Center for Biomedical Research in Network of Bioengineering, Biomaterials and Nanomedicine (CIBER-BBN), 47011 Valladolid, Spain}
\affil[14]{Institute for Research in Mathematics (IMUVA), University of Valladolid, 47011 Valladolid, Spain}
\affil[*]{Address correspondence to: Santiago Cepeda MD., Ph.D.,  \texttt{scepedac@saludcastillayleon.es}}
\affil[$\dag$]{These authors contributed equally to this work.}

\begin{document}
\date{}
\maketitle

\clearpage
\begin{abstract}
Intraoperative ultrasound (ioUS) is a valuable tool in brain tumor surgery due to its versatility, affordability, and seamless integration into the surgical workflow. However, its adoption remains limited, primarily because of the challenges associated with image interpretation and the steep learning curve required for effective use. This study aimed to enhance the interpretability of ioUS images by developing a real-time brain tumor detection system deployable in the operating room.
We collected 2D ioUS images from the Brain Tumor Intraoperative Ultrasound Database (BraTioUS) and the public ReMIND dataset, annotated with expert-refined tumor labels. Using the YOLO11 architecture and its variants, we trained object detection models to identify brain tumors. The dataset included 1,732 images from 192 patients, divided into training, validation, and test sets. Data augmentation expanded the training set to 11,570 images. In the test dataset, YOLO11s achieved the best balance of precision and computational efficiency, with a mAP@50 of 0.95, mAP@50-95 of 0.65, and a processing speed of 34.16 frames per second. The proposed solution was prospectively validated in a cohort of 15 consecutively operated patients diagnosed with brain tumors. Neurosurgeons confirmed its seamless integration into the surgical workflow, with real-time predictions accurately delineating tumor regions.
These findings highlight the potential of real-time object detection algorithms to enhance ioUS-guided brain tumor surgery, addressing key challenges in interpretation and providing a foundation for future development of computer vision-based tools for neuro-oncological surgery.

\end{abstract}

\section{INTRODUCTION}
Gliomas are the most common type of primary brain tumor, characterized by their infiltrative nature, which frequently leads to poor prognosis and significant therapeutic challenges \cite{Ostrom2017}. According to the latest classification of central nervous system tumors by the World Health Organization (WHO) \cite{Louis2021}, gliomas encompass various subtypes, each of which are defined by histopathological and molecular characteristics that directly influence their aggressiveness, management, and prognosis. The surgical treatment of gliomas is particularly challenging due to their frequent location near eloquent brain regions, where preserving critical neurological functions can limit the extent of resection. Maximizing tumor removal is a well-established factor associated with prolonged patient survival, emphasizing the importance of achieving an optimal balance between tumor excision and functional preservation. Consequently, the current standard of care is to achieve what is known as maximum safe resection \cite{Karschnia2021}. To facilitate this goal, surgeons have access to several auxiliary intraoperative tools, including fluorescent agents (e.g., 5-aminolevulinic acid, sodium fluorescein), neuronavigation, direct electrical stimulation, and intraoperative imaging techniques. Among the imaging modalities, intraoperative magnetic resonance imaging (ioMR) and intraoperative ultrasound (ioUS) are particularly noteworthy \cite{Chanbour2022}. While ioMR is considered the gold standard, its high cost and demanding logistical requirements limit its accessibility in many centers \cite{Koukoulithras2024}. In contrast, ioUS offers several advantages, including versatility, low cost, and seamless integration into the surgical workflow \cite{Unsgaard2005,Moiyadi2016}. However, effective use of ioUS involves a steep learning curve and is highly operator dependent. Additionally, interpretation of ioUS images is complicated by artifacts, varying acquisition planes, low contrast between healthy and pathological tissues in certain tumors, a suboptimal signal-to-noise ratio, and a limited field of view \cite{Steno2021,Dixon2022}.
Despite significant advancements in ioUS image quality over the past decades, and numerous studies confirming that its use contributes to achieving maximum safe resection, its limitations have nonetheless hindered widespread adoption in neurosurgical practice \cite{Solheim2010, Cepeda2024, Shetty2021}. Enhancing the interpretability of ioUS images could promote its use among surgeons in training and improve the precise identification of tumor margins. This, in turn, has the potential to optimize image-guided surgery and increase the extent of resection, a critical factor in the effective management of gliomas.
Medical imaging has long driven the development of advanced classification, detection, and segmentation systems. Instance segmentation, for example, has been extensively studied in brain tumors, particularly in magnetic resonance imaging (MRI), which dominates the scientific literature \cite{Xu2024}. In contrast, segmentation in ioUS has received significantly less attention, and studies focused on applying these methods in real-time intraoperative settings are even more scarce \cite{Dorent2024}.
The advent of techniques based on deep learning has revolutionized the interpretation of medical images, facilitating the automatic detection of anomalies without the need for precision at the pixel level \cite{Albuquerque2025}. Object detection has been established as a key solution for real-time analysis in dynamic environments such as agriculture, transportation, education, and health care \cite{Kaur2023}. Among the most prominent algorithms are “You Only Look Once” (YOLO), introduced by Redmon et al. \cite{Redmon2016}, which revolutionized real-time detection by integrating region proposals and classification into a single neural network.
In this work, we propose improving the interpretability of ioUS through real-time tumor detection. Segmentation is considered a secondary objective, as pixel-level precision is not necessary during surgery to delineate tumor margins. Instead, the surgeon determines the limits of resection based on preoperative planning and factors such as tumor location, proximity to eloquent areas, suspected tumor type, and patient-specific characteristics. Consequently, we prioritize detection as the primary objective, utilizing bounding boxes to locate the tumor, improving image interpretation, and correlate it with adjacent brain structures without disrupting visualization or the surgical workflow. Furthermore, accurate detection of residual tumor has the potential to facilitate image-guided surgery and improve resection outcomes.
Our approach is based on training a YOLO11 model using ioUS images from a multicenter dataset and evaluating its real-time implementation in prospective cases. This work introduces several innovative contributions:

\begin{itemize}
    \item \textbf{High-quality multicenter dataset:} Images from the Brain Tumor Intraoperative Ultrasound Database (BraTioUS), which consists of data from six centers from various manufacturers and acquisition protocols, will be used.
    \item \textbf{Glioma subtypes:} The model will be trained with low- and high-grade gliomas, facing the challenge of identifying tumors with imprecise borders, even for experts.
    \item \textbf{Expert refined pseudolabels:} Pseudolabels are generated via a preliminary model \cite{Cepeda2025} and are manually corrected by expert observers.
    \item \textbf{Exclusive use of ioUS:} The model does not depend on images from other modalities, such as MRI, or on generated or coregistered segmentations, facilitating its clinical applicability.
    \item \textbf{Real-time implementation:} The model will be developed specifically for object detection and evaluated in a real surgical environment via low-cost systems.
    \item \textbf{Comparative analysis of YOLO11 variants:} All available variants are evaluated to establish a benchmark in terms of precision and computational efficiency.
\end{itemize}

This study is structured into several sections. 
Section~\ref{sec:literature} provides a comprehensive review of the relevant literature. 
Section~\ref{sec:dataset} outlines the dataset and the process of generating labels. 
Section~\ref{sec:architecture} describes the YOLO11 architecture in detail. 
Section~\ref{sec:evaluation} explains the evaluation criteria and experimental environment. 
Section~\ref{sec:results} presents the experimental results, emphasizing the real-time application. 
Section~\ref{sec:discussion} discusses the findings, compares them with previous studies, and highlights future research directions. 
Finally, Section~\ref{sec:conclusion} concludes the study by summarizing the main contributions.

\section{LITERATURE REVIEW}
\label{sec:literature}
The segmentation of brain tumors, particularly gliomas, has been extensively studied via MRI, which is the predominant modality in most published works \cite{Ali2021, Bakas2018, deVerdier2024}. However, the identification of brain tumors using ioUS imaging has received considerably less attention, despite its increasing interest.
The first relevant study on segmentation in ioUS was published by Ritschel et al. in 2014 \cite{Ritschel2015}, where the authors proposed a pixel-level classification model using contrast-enhanced ultrasound (CEUS) images. A support vector machine (SVM) classifier was trained to differentiate tumor tissue from nontumor tissue in patients with glioblastomas, achieving specificity values of 0.94 and a sensitivity of 0.76. This approach was based on characteristics derived from the temporal analysis of each pixel in the video data, demonstrating segmentations similar to those performed by expert neurosurgeons.
In 2018, Ilunga-Mbuyamba et al. \cite{IlungaMbuyamba2018} developed a patient-specific method using 3D ioUS images and tumor segmentations derived from MRI data through coregistration. In a study with 33 patients, Dice values of 0.77 and 0.76 were reached for glioblastomas when rigid registration and related transformations were used, respectively. Later, Canalini et al. \cite{Canalini2019} introduced a 3D U-Net-based model for ioUS volume segmentation in glioma resections via the public RESECT \cite{Xiao2017} and BITE \cite{Mercier2012} datasets. This work improved the registration between ioUS and MRI via manual segmentation of brain structures such as the sulci and the falx.
In 2020, Carton et al. \cite{Carton2020} implemented a surgical cavity segmentation model based on U-Net, also using the RESECT \cite{Xiao2017} and BITE \cite{Mercier2012} datasets. Its best model, trained with 3D U-Net, reached a median Dice coefficient of 0.88. In 2021, this group expanded its work by incorporating tumors and surgical cavities \cite{Carton2021}, obtaining a Dice coefficient of approximately 0.7 for tumors in preresection images and 0.6 during resection. They recognized limitations such as the small size of the dataset and class imbalance. In the same year, Angel-Raya et al. \cite{AngelRaya2021} evaluated semiautomatic and registry-based methods for segmenting tumors in 3D ioUS data, achieving Dice scores of 0.85 to 0.86 in a set of 66 patients, although they highlighted the dependence on user interaction as a key limitation.
In more recent studies, Dorent et al. \cite{Dorent2024} proposed a patient-specific model based on preoperative MRI and synthetic ioUS images, which achieved Dice scores greater than 0.8 in a limited dataset. On the other hand, Faanes et al. \cite{Faanes2024} developed an automatic model using nnU-Net with data from RESECT-SEG \cite{Behboodi2024}, ReMIND \cite{Juvekar2024} and CuRIOUS-SEG, achieving an average Dice coefficient of 0.62, although they reported lower success rates in small tumors.
Our group has also contributed to this area with a model based on nnU-Net \cite{isensee2021nnunet} trained with images from the multicenter BraTioUS and ReMIND \cite{Juvekar2024} datasets. This model achieved Dice scores of 0.9 in an internal cohort, 0.93 in an independent cohort and 0.65 in RESECT-SEG \cite{Behboodi2024}, indicating good generalizability, although it was conditioned by the selection of a single 2D slice per patient \cite{Cepeda2025}.
Despite advancements in segmentation, none of the aforementioned methods have been specifically developed for real-time object detection or tested in real-life clinical settings. On the other hand, in other areas of medical ultrasound, the detection and segmentation of objects in real time has significantly advanced. For example, Hu et al. \cite{Hu2022} developed a model based on U-Net to segment breast tumors in real time, achieving a Dice coefficient of 0.78 with a rate of 16 FPS. Kumar et al. \cite{Kumar2018} used a Multi U-Net approach, reaching a Dice coefficient of 0.82 with an inference speed between 13 and 55 ms per image. Wei et al. \cite{Wei2024} implemented YOLOv4 for the detection of carotid plaques, achieving an accuracy of 98.5\% and a speed of 39 frames per second (FPS).
Other recent works, such as that of Nurmaini et al. \cite{Nurmaini2024}, used YOLOv8l for the detection of heart defects in pediatric ultrasound videos, achieving a performance of 90.38 FPS with an accuracy of 97.17\%. 
Zhang et al. \cite{Zhang2023} applied a weakly supervised approach to the detection of median nerves, reaching an AP50 of 93.9\%, with a rate of 38.5 FPS. Finally, Ou et al. \cite{Ou2024} proposed RTSeg-Net, a lightweight model for real-time segmentation of fetal heads, which reached a Dice coefficient of 0.85 and a speed of 31.3 FPS.
In summary, while instance segmentation has been the predominant focus in ioUS, real-time object detection represents an unexplored opportunity in the context of brain tumors, highlighting the need to develop robust, efficient and clinically viable systems. This work directly addresses this gap, proposing an innovative approach to real-time tumor detection in ioUS images.

\section{DATASET}
\label{sec:dataset}

\subsection{Study Population}

The main dataset for this study comes from the BraTioUS Consortium (ClinicalTrials.gov Identifier: NCT05062772), which includes contributions from six international institutions: Río Hortega University Hospital, Valladolid, Spain (RHUH); the Tata Memorial Center, Mumbai, India (TMC); the Istituto Neurologico Carlo Besta, Milan, Italy (INCC); the University of Palermo, Italy (UPALER); Le Center Hospitalier Régional Universitaire de Tours, France (CHRUT); and the Massachusetts General Hospital, Boston, USA (MGH).
This dataset is composed of ioUS images obtained from patients who underwent surgery for brain tumors between 2018 and 2023. In this study, only patients diagnosed with gliomas were included, based on the 2021 WHO Classification of Tumors of the Central Nervous System \cite{Louis2021}. Only preresection images in B mode were selected, excluding those cases with different histopathological diagnoses, images of suboptimal quality or artifacts that made their processing and interpretation difficult.
Additionally, images from the public ReMIND \cite{Juvekar2024} dataset were incorporated, applying the same selection criteria. The BraTioUS dataset included a total of 154 patients, 152 of whom met the selection criteria. From the ReMIND dataset, 45 patients were selected from an initial group of 114. In total, the dataset included 197 subjects, of which 128 had native 2D ioUS images and 69 had 3D images. The 3D volumes were processed by extracting slices along the axial plane, corresponding to the third dimension of the data array. Each slice was adjusted to preserve spatial orientation and anatomical consistency, resulting in a series of 2D images for further analysis.
The final dataset consisted of 1,732 intraoperative ultrasound (ioUS) images, as each patient contributing multiple images from their 2D ioUS studies. To evaluate the real-time implementation of the model in the operating room, a prospective study was conducted at the Department of Neurosurgery, Río Hortega University Hospital, between November 2024 and January 2025. Fifteen patients undergoing consecutive craniotomies for brain tumor resections were included in this phase of the study. The applicability of the developed system was qualitatively assessed during these procedures. Table 1 summarizes the complete dataset, while Table 2 provides details of the prospective patient cohort. The use of anonymized data was approved by the Research Ethics Committee (CEIm) of the Río Hortega University Hospital, Valladolid, Spain (approval number 21-PI085). In addition, all patients included in the prospective study signed an informed consent form.

\begin{table}[ht]
\centering
\caption{Demographic Characteristics and Ultrasound Image Acquisition Details Across Centers}
\label{tab:demographics}
\begin{adjustbox}{max width=\textwidth}
\begin{threeparttable}
\begin{tabular}{|l|c|c|c|c|c|c|c|}
\hline
\textbf{Variable}                  & \textbf{RHUH}         & \textbf{ReMIND}       & \textbf{TMC}          & \textbf{CHRUT}        & \textbf{UPALER}       & \textbf{INCC}         & \textbf{MGH}          \\ \hline
\textbf{Total of subjects}         & 58                   & 45                   & 35                   & 29                   & 15                   & 10                   & 5                    \\ \hline
\textbf{Total number of images}    & 738                  & 45                   & 666                  & 177                  & 32                   & 55                   & 19                   \\ \hline
\textbf{Mean Age}                  & 61.66 ± 11.22        & 42.49 ± 15.16        & 47.62 ± 10.72        & 48.16 ± 13.76        & 67.33 ± 12.98        & 58.9 ± 18.77         & NA                   \\ \hline
\textbf{Sex}                       &                      &                      &                      &                      &                      &                      &                      \\ \hline
Male                               & 36 (62.07\%)         & 17 (37.78\%)         & 24 (68.57\%)         & 9 (30\%)             & 5 (33.33\%)          & 6 (60\%)             & NA                   \\ \hline
Female                             & 22 (37.93\%)         & 28 (62.22\%)         & 11 (31.43\%)         & 13 (43.33\%)         & 10 (66.67\%)         & 4 (40\%)             & NA                   \\ \hline
NA                                 & -                    & -                    & -                    & 8 (26.67\%)          & -                    & -                    & NA                   \\ \hline
\textbf{Glioma type}               &                      &                      &                      &                      &                      &                      &                      \\ \hline
Low grade                          & -                    & 11 (24.44\%)         & -                    & 8 (27.59\%)          & -                    & -                    & -                    \\ \hline
High grade                         & 58 (100\%)           & 28 (62.23\%)         & 35 (100\%)           & 29 (72.42\%)         & 10 (66.67\%)         & 10 (100\%)           & 5 (100\%)            \\ \hline
NA                                 & -                    & 6 (13.33\%)          & -                    & -                    & 5 (33.33\%)          & -                    & NA                   \\ \hline
\textbf{US manufacturer}           & Hitachi              & BK                   & BK / Sonowand        & Supersonic           & Esaote               & Esaote               & BK                   \\ \hline
\textbf{Type of probe}             & Curved               & Curved               & Curved               & Linear               & Linear               & Linear               & Curved               \\ \hline
\textbf{Frequency}                 & 4–8 MHz             & 5–13 MHz            & 3–8 MHz             & 4–15 MHz            & 3–11 MHz            & 3–11 MHz            & 5–13 MHz            \\ \hline
\textbf{Acquisition type}          &                      &                      &                      &                      &                      &                      &                      \\ \hline
2D                                 & 58 (100\%)           & -                    & 11 (31.42\%)         & 29 (100\%)           & 15 (100\%)           & 10 (100\%)           & 5 (100\%)            \\ \hline
3D                                 & -                    & 45 (100\%)           & 24 (68.57\%)         & -                    & -                    & -                    & -                    \\ \hline
\end{tabular}
\begin{tablenotes}
\small
\item The values are expressed as standard deviation and percentages as applicable. WHO = World Health Organization. IDH = Isocitrate Dehydrogenase. US = Ultrasound. MHz = Megahertz. NA = not available.
\end{tablenotes}
\end{threeparttable}
\end{adjustbox}
\end{table}

\begin{table}[ht]
\centering
\caption{Clinical and Demographic Characteristics of the Prospective Patient Cohort}
\label{tab:clinical_characteristics}
\begin{adjustbox}{max width=\textwidth}
\begin{tabular}{|c|c|c|c|c|}
\hline
\textbf{Subject ID} & \textbf{Sex} & \textbf{Age} & \textbf{Histology}          & \textbf{Location}             \\ \hline
1                   & Male         & 58           & Glioblastoma                & Right temporal                \\ \hline
2                   & Male         & 44           & Anaplastic astrocytoma      & Left frontal                  \\ \hline
3                   & Female       & 64           & Meningioma Grade 1          & Cerebellopontine angle        \\ \hline
4                   & Male         & 73           & Glioblastoma                & Left parieto-occipital        \\ \hline
5                   & Male         & 63           & Meningioma Grade 2          & Right frontal                 \\ \hline
6                   & Female       & 58           & Meningioma Grade 1          & Left frontoparietal           \\ \hline
7                   & Male         & 33           & Astrocytoma Grade 2         & Right frontal                 \\ \hline
8                   & Male         & 46           & Lung metastasis             & Cerebellar                    \\ \hline
9                   & Female       & 56           & Meningioma Grade 1          & Left frontal                  \\ \hline
10                  & Female       & 60           & Meningioma Grade 1          & Left temporal                 \\ \hline
11                  & Female       & 53           & Glioblastoma                & Left parietal                 \\ \hline
12                  & Male         & 50           & Glioblastoma                & Left temporal                 \\ \hline
13                  & Female       & 55           & Glioblastoma                & Left frontoinsular            \\ \hline
14                  & Female       & 48           & Meningioma Grade 1          & Left frontotemporal           \\ \hline
15                  & Female       & 24           & Astrocytoma Grade 2         & Left frontal                  \\ \hline
\end{tabular}
\end{adjustbox}
\end{table}

\subsection{Ground truth segmentation}
For the initial segmentation process, one representative 2D axial slice was selected for each patient, resulting in a total of 197 images. These slices were used to train a preliminary segmentation model based on the nnU-Net framework \cite{isensee2021nnunet}, previously developed and published by our group. Tumors in these selected slices were manually segmented using ITK-SNAP software (version 4.0.1, \url{http://itksnap.org}), with extensive necrotic or cystic regions excluded from the segmentations. Detailed information about the development and validation of this preliminary model can be found in the corresponding publication \cite{Cepeda2025}.
Using this trained model, we generated pseudo-labels for the remaining 1,535 images in the dataset, which lacked manual segmentations. This semi-automated approach was adopted to overcome the considerable time required for manual annotation, leveraging the robust performance of the preliminary model. The pseudo-labels were subsequently reviewed and refined by two neurosurgeons with expertise in ioUS imaging and neuro-oncology. Any discrepancies were resolved through consensus. Unlike the initial model, these segmentations included the necrotic and cystic regions of the tumors, ensuring that the entire tumor extent was represented.
Both the images and annotations were converted from NIfTI format (Neuroimaging Informatics Technology Initiative) to JPG format (Joint Photographic Experts Group) and TXT format. These conversions were necessary to meet the format requirements of the YOLO11 model. Additionally, tags were generated in both instance segmentation and bounding box formats, ensuring compatibility and flexibility for detection and segmentation processes.

\section{YOLO11 ARCHITECTURE}
\label{sec:architecture}
YOLO architecture divides the image into a grid, predicts bounding boxes and class probabilities for each cell, and enables end-to-end learning \cite{Redmon2016}. The YOLO architecture is composed of three main components: the backbone, the neck and the head. The backbone acts as the primary feature extractor, using convolutional neural networks to transform the original image into a set of feature maps at multiple scales, capturing key structural information. The neck is an intermediate processing stage that adds and refines the characteristics extracted by the backbone, allowing multiple scales to contribute to the detection of objects of various sizes. Finally, the head is the predictor component that generates the final outputs, including the coordinates of the bounding boxes, the classes of the objects and their associated probabilities \cite{Khanam2024}. Figure 1 provides a schematic illustration of the YOLO11 architecture.

\begin{figure}[ht]
    \centering
    \includegraphics[width=\textwidth]{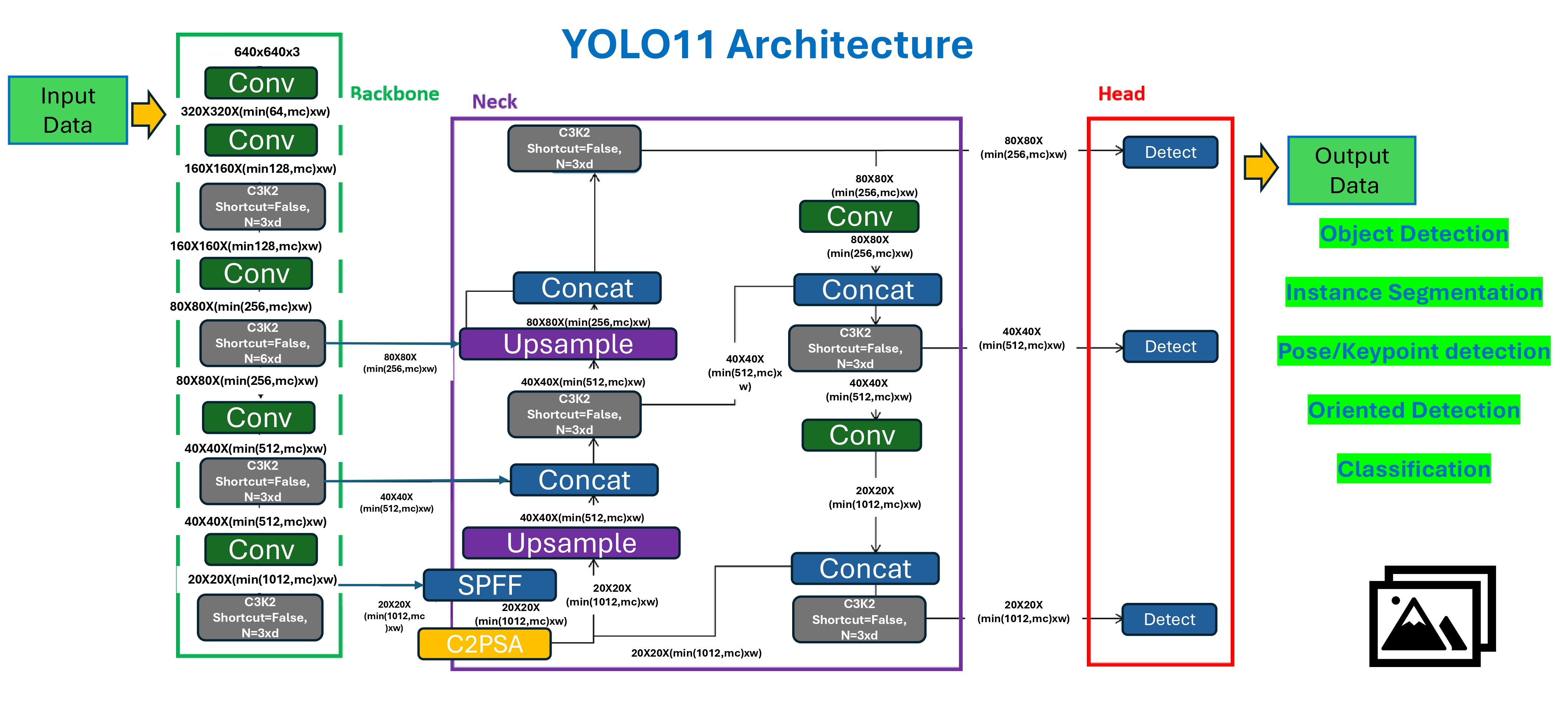} 
    \caption{Schematic representation of YOLO11 architecture. This illustration highlights the key components and flow of the model. Reproduced with permission from the original author \cite{Sapkota2025}.}
    \label{Figure_1} 
\end{figure}

\subsection{YOLO11 enhancements}
Developed by Ultralytics \cite{Ultralytics2023}, YOLO11 incorporates several significant innovations over its predecessors. First, it introduces the C3k2 block (cross-stage partial with a size 2 kernel), which replaces the C2f block from previous versions \cite{Jocher2024}. This block uses two smaller convolutions instead of one large convolution, which improves computational efficiency without sacrificing performance. Second, it maintains the SPPF (Spatial Pyramid Pooling - Fast) component, which allows efficient aggregation of information at different scales, optimizing the capture of relevant characteristics for objects with variable sizes and locations \cite{Jocher2024}. Third, it adds the C2PSA (convolutional block with parallel spatial attention) module, which introduces a parallel spatial attention mechanism, allowing the model to focus efficiently on specific regions of the image and improving precision for objects with varied positions and sizes. In addition, it includes the CBS (Convolutional-Batch-Norm-SiLU) layers, which refine the characteristic maps by extracting relevant information, normalizing the data flow and using the SiLU (Sigmoid Linear Unit) activation function, known for combining precision and numerical stability. Finally, the final Conv2D layers reduce the refined features to the required number of outputs, generating the bounding box coordinates and class predictions \cite{Khanam2024}.

\subsection{Additional Capabilities of YOLO11}
In addition to object detection, YOLO11 supports multiple computer vision tasks. These include instance segmentation, which identifies the precise contours of each object within an image; image classification, which categorizes entire images into predefined classes; pose estimation, which determines the position and orientation of key parts of objects; oriented object detection, which detects objects while considering their orientation in space; and object tracking, which continuously tracks objects in video sequences \cite{Ultralytics2023}.
These capabilities make YOLO11 a versatile, efficient and highly suitable architecture for real-time tasks in critical applications, such as the detection of brain tumors via intraoperative ultrasound. The combination of precision, computational efficiency and real-time processing capacity makes YOLO11 an ideal tool for dynamic surgical environments, optimizing visual support during procedures.

\subsection{Data from pretrained models}
 The YOLO11 models are initially trained on the Common Objects in Context (COCO) dataset \cite{Lin2015}, which is widely recognized in computer vision tasks. This dataset includes approximately 330,000 images, of which 200,000 are annotated for tasks such as object detection, image segmentation, and captioning. COCO covers 80 categories of objects, from everyday items, such as cars, animals and food, to specific items, such as bags and sports equipment. Its structure is organized into three main subsets: Train2017, composed of 118,000 images used for training; Val2017, which contains 5,000 images for validation during training; and Test2017, with 20,000 images intended for benchmarking. The pretrained YOLO11 models, developed by Ultralytics \cite{Ultralytics2023}, are available for download in various versions: YOLO11n, YOLO11 s, YOLO11m, YOLO11l and YOLO11x. These variants present differences in size, average precision and inference speed, allowing the selection of the most appropriate one according to the specific requirements of the task and the available computational resources.

 \section{EVALUATION CRITERIA AND EXPERIMENTAL ENVIRONMENT}
 \label{sec:evaluation}
This section presents the metrics used to evaluate the model and measure the computational resources consumed. For the evaluation of the model, standard metrics were employed, including precision, F1 score, recall, mAP@50 (mean Average Precision at 50\% Intersection over Union), and mAP@50-95 (mean Average Precision at Intersection over Union thresholds ranging from 50\% to 95\%), which are commonly used in detection tasks. These metrics enable the evaluation of the model's ability to detect objects with both precision and consistency. In the instance segmentation task, specific metrics such as the Dice similarity coefficient (DSC) and the Jaccard index or intersection over union (IoU) were added, which are essential for measuring the overlap between the predictions and the ground truth.
In terms of computational resources, key parameters such as model size, memory consumption, number of parameters, FLOPS (floating point operations per second), MADD (multiplication and addition operations), FPS, and latency were analyzed to evaluate the model's viability for real-time applications.
In addition, the characteristics of the experimental environment are detailed, including the hardware used, the operating system, the configuration and the hyperparameters of the model during training. Finally, the steps implemented for data preprocessing; the division of the dataset into training, validation and test sets; and the data augmentation techniques used to improve the generalizability of the model are described.

\subsection{Model performance evaluation metrics}
To evaluate the model's performance, various metrics were employed to provide a comprehensive assessment of its capabilities. In the main task of object detection, precision was used, defined as the proportion of correct positive predictions with respect to the total number of positive predictions made by the model. This metric is key to determining the ability of the model to minimize false positives, ensuring that detections are relevant and accurate.
\[
\text{Precision} = \frac{\text{True Positives (TP)}}{\text{True Positives (TP)} + \text{False Positives (FP)}}
\]
TP: Number of positive cases correctly identified.
FP: Number of negative cases incorrectly identified as positive.
Sensitivity measures the ability of a model to correctly identify positive cases among all actual positive cases.
\[
\text{Sensitivity} = \frac{\text{True Positives (TP)}}{\text{True Positives (TP)} + \text{False Negatives (FN)}}
\]
FN: Number of positive cases that the model did not identify correctly.
The F1 score is the harmonic mean between precision and sensitivity (recall). It is particularly useful in scenarios with class imbalance, as it balances the contributions of both metrics into a single comprehensive value.
\[
F_1 = 2 \cdot \frac{\text{Precision} \cdot \text{Sensitivity}}{\text{Precision} + \text{Sensitivity}}
\]
The mAP is a metric used to evaluate the performance of object detection models. It combines the precision and recall metrics into a single value, calculating the precision average for different IoU thresholds.
 The mAP@50 measures the average precision considering a fixed threshold of the IoU equal to 0.50. This means that a predicted bounding box is considered correct if it has at least 50\% overlap with the actual bounding box.
\[
AP = \int_{0}^{1} P(R) \, dR
\]
Where P (R) is the precision curve as a function of recall. The calculation of the mAP@50 is simply the average of the accuracies (AP) for all classes at the threshold IoU = 0.50.
The mAP@50-95 extends the mAP calculation by considering multiple values of the IoU, from 0.50 to 0.95, incremented in steps of 0.05. It is more demanding than mAP@50 because it evaluates the performance of the model at different levels of overlap.
\[
\text{mAP@50:95} = \frac{1}{10} \sum_{\text{IoU}=0.50}^{0.95} AP(\text{IoU})
\]
The AP is calculated for IoU = 0.50, 0.55, 0.60, …, 0.95, and the result is the average.
For the second task of this work, instance segmentation, we add the overlap metrics:
The Jaccard index measures the ratio between the area of intersection and the area of union of two sets and is used to evaluate object detection or segmentation.
\[
\text{IoU} = \frac{\lvert \text{Intersection} \rvert}{\lvert \text{Union} \rvert}
\]
Intersection: The common area between the prediction and ground truth regions.
Union: The sum of the prediction and ground truth areas minus the intersection.
The DCS measures the similarity between two sets and is commonly used to evaluate the overlap between segmented regions (prediction) and real regions (ground truth).
\[
\text{DSC} = \frac{2 \cdot \lvert \text{Intersection} \rvert}{\lvert \text{Prediction} \rvert + \lvert \text{Ground Truth} \rvert}
\]
Intersection: The overlapping area between the prediction and ground truth regions.
 Prediction: The total area of the predicted region.
 Ground Truth: The total area of the actual region.

\subsection{Computational efficiency measures}
To evaluate the computational efficiency of the developed detection model, various metrics were analyzed to characterize its performance in terms of storage, resource consumption and processing capacity.
The size of the model (size) refers to the space required for its storage on disk, which is generally expressed in megabytes (MB), a crucial factor for systems with storage restrictions. Memory (memory) measures the amount of RAM or VRAM consumed during inference, which is particularly important for devices with limited resources.
The number of parameters (parameters) reflects the number of trainable weights in the model, a direct indicator of its complexity and ability to learn detailed representations. To measure the computational load, metrics such as floating-point operations per second (FLOPS) and multiplication and addition operations (MADD) were used, which allow evaluation of the efficiency of the model in terms of mathematical processing.
Additionally, metrics related to real-time performance, such as FPS and latency (latency), were analyzed. The FPS measures the number of images processed in one second, while the latency corresponds to the time required to process a single image, expressed in milliseconds (ms). These metrics are essential for determining the feasibility of the model in applications where the response time is critical, such as in surgical environments.

\subsection{Experimental environment}
The model was trained on a Linux Ubuntu 24.04.1 LTS operating system, using Python 3.10 as the programming language. To implement and train the model, the deep learning framework PyTorch 2.5.1 was used, in combination with CUDA 12.4 and CuDNN 9.5.1, taking advantage of GPU acceleration. The hardware used included an AMD Ryzen 9 processor and an RTX 4080 GPU with 16 GB of memory and 32 GB of RAM, ensuring a high-performance environment for intensive training tasks.
For the evaluation on the test dataset, prospective evaluation, and real-time implementation in the operating room, a laptop computer with Windows 10 as the operating system and the following software configuration was used: Python 3.10, PyTorch 2.5.1, CUDA 12.1 and CuDNN 9.1.0. In this case, the hardware consisted of an Intel Core i7 processor, an RTX 3070 GPU with 8 GB of memory and 32 GB of RAM, which provided a suitable environment for testing in clinical conditions and real-time applications.

\subsection{Intraoperative acquisition technique and inference hardware setup}
For the prospective evaluation, all patients included in the study underwent craniotomy. A Hitachi Noblus ultrasound system with a C42 microconvex probe operating in a frequency range of 4-8 MHz was used for intraoperative image acquisition. The probe was covered with sterile sheets, and a minimal amount of conductive gel was applied to avoid interference. The B-mode images were initially captured in different planes for reference, and ultrasound studies were performed by neurosurgeons [SC, RS, OE-S.]. In each case, at least two acquisitions were made: one before resection and another after resection, with additional acquisitions according to the needs of the procedure.
For the implementation of the model in prospective cases, the video capture device USB3HDCAP from StarTech.com, an external device that allows the recording of content at high definition (1080p at 60 fps) through a USB 3.0 connection, was used. As input, the DVI port compatible with the Noblus ultrasound system was used. This capturer uses the H.264 codec for video encoding and is compatible with Microsoft® DirectShow, which facilitates its integration with video capture and editing software.
An optimized inference script was executed on the laptop within the operating room to load the trained model and enable real-time operation during prospective cases. The ultrasound scanner and laptop screens were positioned side-by-side, facilitating simultaneous assessment of the model's detection and segmentation capabilities and its integration into the surgical workflow. This setup allowed for a qualitative evaluation of the model's applicability in a real clinical setting, ensuring its implementation did not disrupt ongoing surgical procedures.

\subsection{YOLO11 hyperparameter values}
During the training of the YOLO11 model, using the Ultralytics implementation, default hyperparameters designed to optimize performance in object detection tasks were used. The training was carried out for 500 epochs, using an initial learning rate (\texttt{lr0}) of 0.01, which progressively decreased according to a final factor (\texttt{lrf}) of 0.01. The batch size was established at 16 images, with an input size of 640 pixels, while a momentum of 0.937 was applied to stabilize the updates of the weights.
To avoid overfitting, a weight decay (\texttt{weight\_decay}) of 0.0005 was used. In addition, a warmup strategy was implemented during the first 3 seasons, gradually adjusting the learning rate and momentum. The initial \texttt{warmup\_momentum} was 0.8, and \texttt{warmup\_bias\_lr} started at 0.1, favoring a smooth transition to full training.
The model uses 3 anchors per feature level, which optimizes the detection of objects at various scales. Among the activation functions used are leaky ReLU, which introduces nonlinearity and improves the ability of the model to capture complex relationships in the data.
The default inference settings include a confidence threshold of 0.25, which defines the minimum level of confidence required to accept a detection, and an IoU threshold of 0.7, which is used in nonmaximum suppression (NMS) to reduce redundant detections.
YOLO11 pretrained weights were used in all its variants for both object detection tasks and, for instance, segmentation. For the detection of objects, the models yolo11n.pt, yolo11 s.pt, yolo11m.pt, yolo11l.pt and yolo11x.pt were used, which cover different levels of complexity and efficiency. For the segmentation of instances, the variants yolo11n-seg.pt, yolo11 s-seg.pt, yolo11m-seg.pt, yolo11l-seg.pt and yolo11x-seg.pt were used, adapted to meet specific needs for precision and computational consumption.

\subsection{Data split and data augmentation}
The dataset of 1,732 images was randomly divided into three subsets: training, validation, and testing, at a ratio of 70:10:20, with stratification by subject. This approach ensured that all the images corresponding to the same subject remained in a single subset, preventing related images from appearing in several strata and generating biases in the results.
For preprocessing and data augmentation, the Roboflow web platform (\url{https://roboflow.com}) was used \cite{Dwyer2024}. During the preprocessing stage, the images were self-oriented to ensure uniform alignment and were resized to 640x640 pixels by stretching. Various data augmentation techniques were subsequently applied, generating up to 10 transformed versions for each image of the training set to increase the diversity of the dataset. The transformations performed included horizontal and vertical flips, 90° clockwise, counterclockwise and upside-down rotations, and clipping with a zoom that ranged from a minimum of 0\% to a maximum of 20\%. In addition, random rotations were made between -15° and + 15°, horizontal and vertical shears of up to ± 10°, and saturation and brightness adjustments that varied between -25\% and + 25\% and between -15\% and + 15\%, respectively, were used. Blurs of up to 2.5 pixels and the addition of random noise were also applied to a maximum of 0.1\% of the pixels.
These data augmentation techniques expanded the training set to a total of 11,570 images, increasing the model's ability to generalize to unseen variations in training. On the other hand, the validation (n = 216) and test (n = 359) sets were not subjected to any data augmentation technique, ensuring that the evaluation accurately reflected the performance of the model on completely new data.
After preprocessing and expansion, the dataset structure was adapted to the format required by YOLO11 and later downloaded from Roboflow for local use. This pipeline of preprocessing and data augmentation was applied consistently for both the detection and segmentation tasks, ensuring uniformity in the implementations and evaluations carried out (Figure 2).

\begin{figure}[ht]
    \centering
    \includegraphics[width=0.9\textwidth]{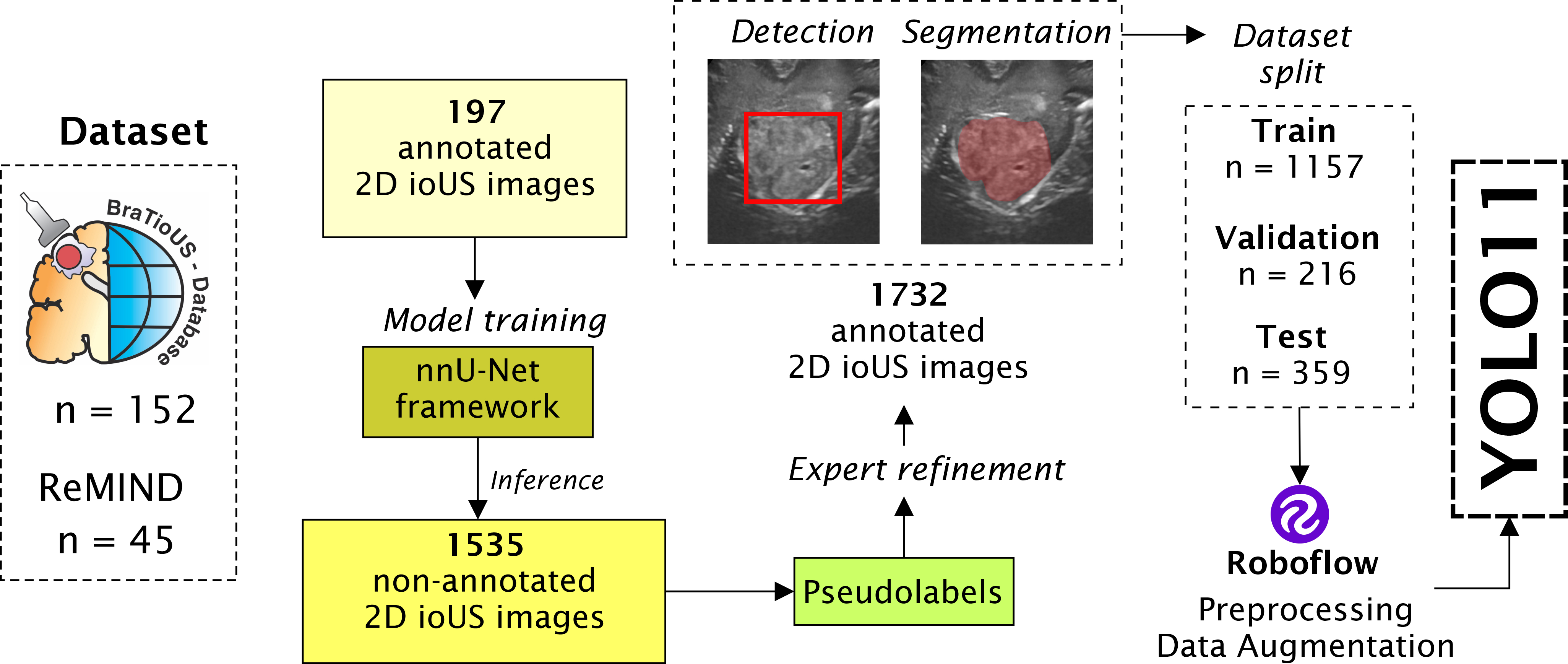}
    \caption{Schematic representation of the workflow data preparation}
    \label{Figure_2}
\end{figure}

\section{EXPERIMENTAL RESULTS}
\label{sec:results}
The primary objectives of this work were to develop an object detection model for brain tumors via intraoperative ultrasound imaging and to evaluate its implementation in real time in a real surgical environment via the trained model. As a secondary objective, an instance segmentation model was developed to complement the analysis. In addition, an exhaustive comparison was made of the results obtained when all the YOLO11 variants were used together with the weights of the pretrained models to determine the optimal balance between the precision and consumption of computational resources, which are fundamental aspects that guarantee their applicability in clinical settings.

\subsection{Object detection task}
Table 3 summarizes the detection metrics of YOLO11 variants, while Table 4 compares their computational requirements and real-time performance. The different variants of the YOLO11 model balance size, efficiency and precision in object detection tasks. The 'n' version is the lightest, with a size of 5.23 MB and 2.59 million parameters, reaching a mAP@50 of 0.94 and a mAP@50-95 of 0.68, with a latency of 30.3 ms and a speed of 25.98 FPS. The 's' version increases the number of parameters to 9.43 million and reduces the latency to 24.9 ms, maintaining a mAP@50 of 0.95 and an mAP@50-95 of 0.65. The 'm' and 'l' versions progressively increase in size and complexity, with 20.05 and 25.31 million parameters, respectively, maintaining mAP@50 values of 0.94 and 0.95, and mAP@50-95 values of 0.66 and 0.69, with latencies of 27.8 ms and 32.9 ms. The 'x' version, the most robust, has 56.87 million parameters, an mAP@50 of 0.93 and an mAP@50-95 of 0.66, with a latency of 42.8 ms and a speed of 21.14 FPS. In terms of precision, all the variants maintain high levels, with recall and precision values above 0.93 and an F1 score close to 0.95. These results show that although the more advanced versions offer improvements in certain aspects, the lighter versions provide remarkable performance with lower computational requirements, allowing adequate selection according to the specific needs of the application. Figure 3 presents several examples of predictions from the test cohort.

\begin{table}[ht]
\centering
\caption{Evaluation Metrics for YOLO11 Variants on Detection Performance}
\label{tab:yolo11_evaluation}
\begin{adjustbox}{max width=\textwidth}
\begin{tabular}{|c|c|c|c|c|c|}
\hline
\textbf{YOLO11 Variant} & \textbf{mAP@50}           & \textbf{mAP@50-95}        & \textbf{Recall}            & \textbf{Precision}         & \textbf{F1}                \\ \hline
n                       & 0.94 (0.92, 0.97)        & 0.68 (0.53, 0.86)        & 0.99 (0.96, 1.00)        & 0.94 (0.92, 0.97)        & 0.96 (0.93, 0.98)         \\ \hline
s                       & 0.95 (0.93, 0.97)        & 0.65 (0.49, 0.82)        & 0.97 (0.94, 1.00)        & 0.94 (0.91, 0.96)        & 0.95 (0.93, 0.97)         \\ \hline
m                       & 0.94 (0.92, 0.97)        & 0.66 (0.50, 0.84)        & 0.96 (0.94, 0.99)        & 0.94 (0.92, 0.97)        & 0.95 (0.92, 0.97)         \\ \hline
l                       & 0.95 (0.93, 0.97)        & 0.69 (0.54, 0.87)        & 0.98 (0.95, 1.01)        & 0.94 (0.91, 0.96)        & 0.95 (0.93, 0.97)         \\ \hline
x                       & 0.93 (0.91, 0.96)        & 0.66 (0.52, 0.84)        & 0.96 (0.93, 0.99)        & 0.93 (0.91, 0.96)        & 0.94 (0.91, 0.97)         \\ \hline
\end{tabular}
\end{adjustbox}
\begin{tablenotes}
\small
\item \textbf{mAP@50}: Mean Average Precision at 50\% Intersection over Union (IoU). 
\item \textbf{mAP@50-95}: Mean Average Precision averaged over IoU thresholds from 50\% to 95\%.
\item The values are expressed as medians and 95\% confidence intervals (CI).
\end{tablenotes}
\end{table}

\begin{table}[ht]
\centering
\caption{Comparison of YOLO Variants Based on Model Size, Computational Requirements, and Real-Time Performance for Object Detection Task}
\label{tab:yolo_variants_comparison}
\begin{adjustbox}{max width=\textwidth}
\begin{tabular}{|c|c|c|c|c|c|c|c|}
\hline
\textbf{YOLO11 Variant} & \textbf{Size (MB)} & \textbf{Memory (MB)} & \textbf{Parameters (millions)} & \textbf{FLOPS (G)} & \textbf{MADD (G)} & \textbf{FPS} & \textbf{Latency (s)} \\ \hline
n                       & 5.23              & 1269.07             & 2.59                          & 3.22               & 1.61              & 25.98        & 0.0303               \\ \hline
s                       & 18.3              & 1255.55             & 9.43                          & 10.77              & 5.39              & 34.16        & 0.0249               \\ \hline
m                       & 38.65             & 1458.39             & 20.05                         & 34.09              & 17.05             & 31.13        & 0.0278               \\ \hline
l                       & 48.83             & 1485.83             & 25.31                         & 43.64              & 21.82             & 26.84        & 0.0329               \\ \hline
x                       & 109.11            & 1700.52             & 56.87                         & 97.72              & 48.86             & 21.14        & 0.0428               \\ \hline
\end{tabular}
\end{adjustbox}
\begin{tablenotes}
\small
\item \textbf{MB}: Megabytes. \textbf{FLOPS}: Floating Point Operations Per Second. 
\item \textbf{MADD}: Multiplication and Addition Operations. \textbf{FPS}: Frames Per Second. \textbf{S}: seconds.
\end{tablenotes}
\end{table}

Table 5 presents the evaluation results of the models in detecting objects after classifying tumor labels by size. Tumor size classification was performed by dividing the bounding box areas into three categories: small (S), medium (M), and large (L). The thresholds for these categories were defined using the 33.3\% and 66.6\% quantiles, with areas exceeding the 66.6\% quantile classified as large. This allowed for a detailed assessment of the model's performance in relation to object dimensions. This approach ensures that the analysis captures the model's capability to address specific detection challenges associated with varying object sizes. The results show that the different variants of YOLO11 have variable performance in the detection of objects according to their size. The average mean precision (mAP@50-95) for small labels (S) ranged between 0.56 and 0.62, with YOLO11n and YOLO11l being the most accurate. For medium labels (M), the variants reached mAP@50-95 values between 0.70 and 0.75, highlighting that YOLO11l had the best performance. For large labels (L), the results ranged between 0.66 and 0.71, with YOLO11m standing out slightly.

\begin{table}[ht]
\centering
\caption{Object Detection Performance of YOLO Variants Across Different Object Sizes}
\label{tab:yolo_performance_sizes}
\begin{adjustbox}{max width=\textwidth}
\begin{tabular}{|c|c|c|c|}
\hline
\textbf{YOLO Variant} & \textbf{mAP@50-95-S}      & \textbf{mAP@50-95-M}      & \textbf{mAP@50-95-L}      \\ \hline
n                     & 0.62 (0.56, 0.68)        & 0.74 (0.70, 0.78)        & 0.69 (0.64, 0.74)        \\ \hline
s                     & 0.59 (0.53, 0.65)        & 0.71 (0.68, 0.75)        & 0.66 (0.61, 0.71)        \\ \hline
m                     & 0.58 (0.52, 0.64)        & 0.70 (0.67, 0.74)        & 0.71 (0.67, 0.76)        \\ \hline
l                     & 0.62 (0.56, 0.68)        & 0.75 (0.72, 0.79)        & 0.69 (0.64, 0.73)        \\ \hline
x                     & 0.56 (0.50, 0.62)        & 0.74 (0.70, 0.78)        & 0.70 (0.66, 0.75)        \\ \hline
\end{tabular}
\end{adjustbox}
\begin{tablenotes}
\small
\item \textbf{mAP@50}: Mean Average Precision at 50\% Intersection over Union (IoU).
\item \textbf{mAP@50-95}: Mean Average Precision averaged over IoU thresholds from 50\% to 95\%.
\item The values are expressed as medians and 95\% confidence intervals (CI).
\end{tablenotes}
\end{table}

\begin{figure}[H]
    \centering
    \includegraphics[width=0.5\textwidth]{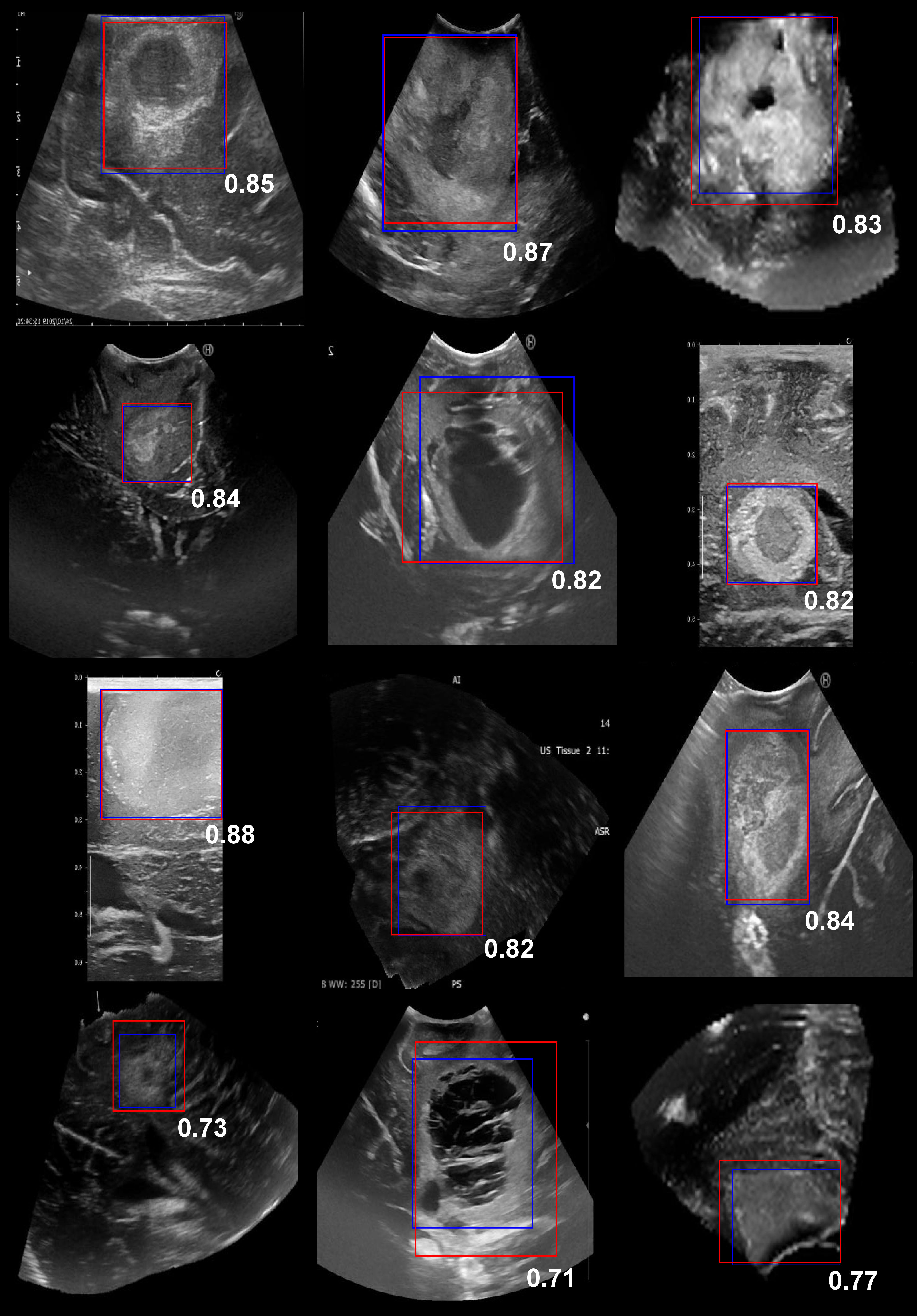}
    \caption{Several examples of tumor detection predictions in the test group. The blue bounding box represents the ground truth (GT), while the red box indicates the model's prediction. The accompanying value represents the confidence score assigned by the model (YOLO11s)}
    \label{Figure_3}
\end{figure}

\subsection{Instance segmentation task}
Table 6 summarizes the segmentation metrics of YOLO11 variants, while Table 7 compares their computational requirements and real-time performance.  There is a balance between performance and computational efficiency. The lighter variants, such as YOLO11n, have a reduced size (5.75 MB), less memory used (1288.02 MB) and 2.83 million parameters, with outstanding performance in metrics such as Dice (0.87) and IoU (0.81). The average precision (mAP@50 and mAP@50-95) of this version reached 0.94 and 0.74, respectively, with latency times of 29.6 ms and 18.87 FPS. On the other hand, the more robust versions, such as YOLO11x, showed higher computational requirements, with 62 million parameters and a consumption of 1321.8 MB of memory, but with similar performances in terms of key metrics: Dice (0.88), IoU (0.82), mAP@50 (0.93) and mAP@50-95 (0.76), although with higher latency (43.5 ms) and lower speed (14.72 FPS). Intermediate versions, such as YOLO11 s and YOLO11m, provide an optimal balance between precision and efficiency, with consistent metrics (mAP@50~0.94, mAP@50-95~0.75) and moderate latencies (\~26 ms). These results highlight that although the more advanced variants offer higher capacities, the lighter versions maintain competitive performance with lower hardware requirements, allowing them to adapt to different scenarios and applications. Figure 4 presents examples of predicted segmentations from the test cohort.

\begin{table}[ht]
\centering
\caption{Evaluation Metrics for YOLO11 Variants on Segmentation Performance}
\label{tab:yolo11_segmentation}
\begin{adjustbox}{max width=\textwidth}
\begin{tabular}{|c|c|c|c|c|c|c|c|}
\hline
\textbf{YOLO11 Variant} & \textbf{DCS}             & \textbf{IoU}             & \textbf{mAP@50}         & \textbf{mAP@50-95}      & \textbf{Recall}          & \textbf{Precision}       & \textbf{F1}             \\ \hline
n                       & 0.87 (0.85, 0.89)        & 0.81 (0.78, 0.83)        & 0.94 (0.91, 0.96)       & 0.74 (0.47, 0.86)       & 0.96 (0.93, 0.97)        & 0.95 (0.93, 0.97)        & 0.95 (0.93, 0.97)       \\ \hline
s                       & 0.88 (0.86, 0.90)        & 0.82 (0.80, 0.84)        & 0.94 (0.91, 0.96)       & 0.76 (0.49, 0.88)       & 0.96 (0.93, 0.97)        & 0.95 (0.93, 0.97)        & 0.95 (0.93, 0.97)       \\ \hline
m                       & 0.88 (0.86, 0.90)        & 0.82 (0.79, 0.84)        & 0.93 (0.90, 0.95)       & 0.74 (0.49, 0.86)       & 0.96 (0.94, 0.98)        & 0.94 (0.91, 0.96)        & 0.94 (0.92, 0.96)       \\ \hline
l                       & 0.87 (0.85, 0.89)        & 0.81 (0.78, 0.83)        & 0.94 (0.91, 0.96)       & 0.75 (0.47, 0.88)       & 0.95 (0.92, 0.97)        & 0.95 (0.92, 0.97)        & 0.94 (0.91, 0.96)       \\ \hline
x                       & 0.88 (0.85, 0.90)        & 0.82 (0.79, 0.84)        & 0.93 (0.91, 0.95)       & 0.76 (0.53, 0.88)       & 0.95 (0.93, 0.97)        & 0.95 (0.92, 0.96)        & 0.94 (0.92, 0.96)       \\ \hline
\end{tabular}
\end{adjustbox}
\begin{tablenotes}
\small
\item \textbf{DCS}: Dice similarity coefficient. \textbf{IoU}: Intersection over union.
\item \textbf{mAP@50}: Mean Average Precision at 50\% Intersection over Union (IoU).
\item \textbf{mAP@50-95}: Mean Average Precision averaged over IoU thresholds from 50\% to 95\%.
\item The values are expressed as medians and 95\% confidence intervals (CI).
\end{tablenotes}
\end{table}

\begin{table}[ht]
\centering
\caption{Comparison of YOLO Variants Based on Model Size, Computational Requirements, and Real-Time Performance for Instance Segmentation Task}
\label{tab:yolo_segmentation_comparison}
\begin{adjustbox}{max width=\textwidth}
\begin{tabular}{|c|c|c|c|c|c|c|c|}
\hline
\textbf{YOLO11 Variant} & \textbf{Size (MB)} & \textbf{Memory (MB)} & \textbf{Parameters (millions)} & \textbf{FLOPS (G)} & \textbf{MADD (G)} & \textbf{FPS} & \textbf{Latency (s)} \\ \hline
n                       & 5.75              & 1288.02             & 2.83                          & 5.1                & 2.55              & 18.87        & 0.0296               \\ \hline
s                       & 19.58             & 1296.02             & 10.07                         & 17.65              & 8.83              & 20.79        & 0.0244               \\ \hline
m                       & 43.08             & 1316.9              & 22.34                         & 61.48              & 30.74             & 19.53        & 0.0265               \\ \hline
l                       & 53.27             & 1306.66             & 27.59                         & 70.94              & 35.47             & 17.98        & 0.0326               \\ \hline
x                       & 119.02            & 1321.8              & 62                            & 159.26             & 79.63             & 14.72        & 0.0435               \\ \hline
\end{tabular}
\end{adjustbox}
\begin{tablenotes}
\small
\item \textbf{MB}: Megabytes. \textbf{FLOPS}: Floating Point Operations Per Second.
\item \textbf{MADD}: Multiplication and Addition Operations. \textbf{FPS}: Frames Per Second. \textbf{S}: seconds.
\end{tablenotes}
\end{table}

Table 8 shows the results of the evaluation of the segmentation evaluated by label size. For small objects, the Dice index ranged between 0.77 and 0.80, while the IoU ranged between 0.70 and 0.74, with mAP@50-95 values ranging from 0.65-0.69. For medium-sized objects, the performance was notably higher, with Dice values between 0.91 and 0.93, IoUs between 0.85 and 0.88, and mAP@50-95 values between 0.74 and 0.80, highlighting the 's' and 'x' versions as the most accurate in this category. For large objects, the Dice and IoU indices remained high, ranging from 0.92--0.93 and 0.86--0.87, respectively, whereas the mAP@50--95 varied between 0.78 and 0.79. This analysis shows that the lighter variants (e.g., 'n') have competitive performance in all categories, whereas the more complex versions (e.g., 'x') present a slight advantage in precision for medium and large objects. This breakdown by size makes it possible to evaluate the capacity of the model to handle different detection and segmentation scales.

\begin{table}[ht]
\centering
\caption{Instance Segmentation Performance of YOLO11 Variants Across Different Object Sizes}
\label{tab:instance_segmentation_performance}
\begin{threeparttable}
\begin{adjustbox}{max width=\textwidth}
\begin{tabular}{|c|c|c|c|c|c|c|c|c|c|}
\hline
\textbf{YOLO11 variant} & \textbf{DSC-S} & \textbf{IoU-S} & \textbf{mAP@50-95-S} & \textbf{DSC-M} & \textbf{IoU-M} & \textbf{mAP@50-95-M} & \textbf{DSC-L} & \textbf{IoU-L} & \textbf{mAP@50-95-L} \\ \hline
\textbf{n} & 0.78 (0.72, 0.84) & 0.71 (0.65, 0.76) & 0.65 (0.59, 0.70) & 0.92 (0.91, 0.93) & 0.85 (0.83, 0.87) & 0.75 (0.72, 0.78) & 0.93 (0.92, 0.94) & 0.86 (0.84, 0.89) & 0.79 (0.76, 0.83) \\ \hline
\textbf{s} & 0.78 (0.72, 0.84) & 0.71 (0.65, 0.77) & 0.67 (0.61, 0.73) & 0.93 (0.93, 0.94) & 0.88 (0.87, 0.89) & 0.80 (0.78, 0.83) & 0.93 (0.92, 0.94) & 0.87 (0.85, 0.89) & 0.79 (0.75, 0.82) \\ \hline
\textbf{m} & 0.80 (0.74, 0.86) & 0.74 (0.69, 0.80) & 0.68 (0.63, 0.74) & 0.91 (0.90, 0.93) & 0.85 (0.83, 0.87) & 0.74 (0.71, 0.77) & 0.93 (0.92, 0.94) & 0.87 (0.85, 0.90) & 0.78 (0.74, 0.82) \\ \hline
\textbf{l} & 0.77 (0.71, 0.83) & 0.70 (0.64, 0.76) & 0.66 (0.61, 0.72) & 0.92 (0.91, 0.93) & 0.86 (0.85, 0.88) & 0.77 (0.74, 0.80) & 0.93 (0.92, 0.94) & 0.86 (0.83, 0.89) & 0.78 (0.75, 0.82) \\ \hline
\textbf{x} & 0.79 (0.73, 0.85) & 0.73 (0.67, 0.79) & 0.69 (0.63, 0.74) & 0.93 (0.92, 0.93) & 0.86 (0.85, 0.88) & 0.77 (0.74, 0.80) & 0.92 (0.90, 0.95) & 0.86 (0.83, 0.89) & 0.79 (0.75, 0.83) \\ \hline
\end{tabular}
\end{adjustbox}
\begin{tablenotes}
\small
\item \textbf{DSC-S}, \textbf{IoU-S}, \textbf{mAP@50-95-S}: Metrics for small objects. 
\item \textbf{DSC-M}, \textbf{IoU-M}, \textbf{mAP@50-95-M}: Metrics for medium objects. 
\item \textbf{DSC-L}, \textbf{IoU-L}, \textbf{mAP@50-95-L}: Metrics for large objects.
\item mAP@50: Mean Average Precision at 50\% Intersection over Union (IoU). 
\item mAP@50-95: Mean Average Precision averaged over IoU thresholds from 50\% to 95\%. 
\item Values are medians with 95\% confidence intervals.
\end{tablenotes}
\end{threeparttable}
\end{table}

\begin{figure}[H]
    \centering
    \includegraphics[width=0.5\textwidth]{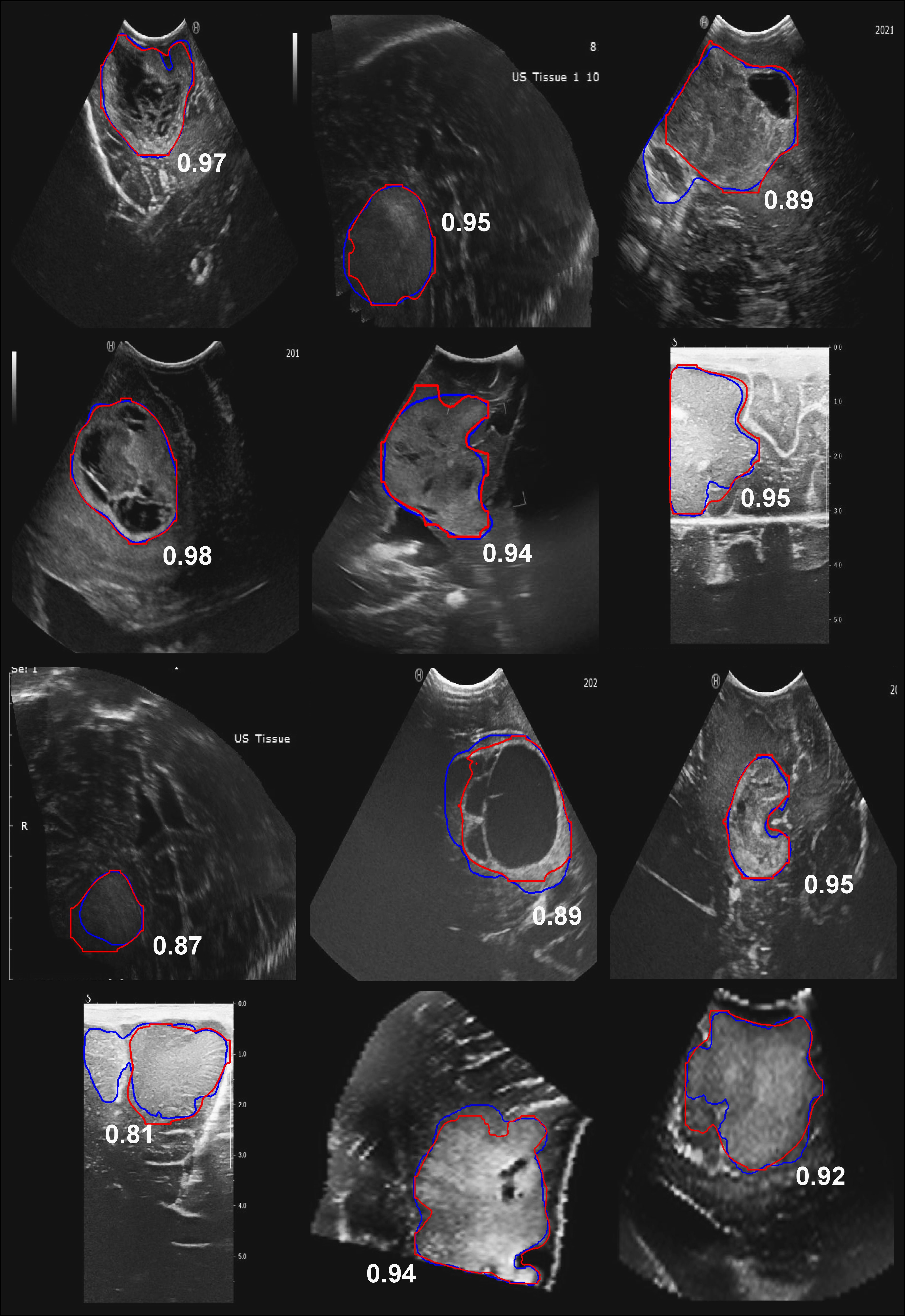}
    \caption{Several examples of tumor segmentation predictions in the test group. The blue region represents the ground truth (GT), while the red region indicates the model's prediction (YOLO11s). The accompanying value represents the Dice coefficient.}
    \label{Figure_4}
\end{figure}

\subsection{Implementation of the model in the operating room}
The characteristics of the fifteen patients for whom the model was implemented during surgery are summarized in Table 2. In the present study, the YOLO11s model was selected for implementation in the operating room because of its optimal balance between precision, speed, and computational efficiency, which are essential characteristics for real-time analysis of brain tumors via intraoperative ultrasound. This model, with a size of 18.3 MB and 9.43 million parameters, presents a latency of 24.9 ms and a speed of 34.16 FPS, which guarantees continuous flow without interruptions during the surgical procedure. In addition, it maintains high levels of precision (mAP@50 of 0.95 and mAP@50-95 of 0.65), comparable to more complex models but with lower hardware requirements. These properties make it ideal for the configuration of the system used.
The results of the model were qualitatively evaluated by the surgeons who participated in the surgical interventions (RS, SC, OE-S). Figure 5 illustrates the operating room setup, including the integration of the object detection model and ultrasound system during the procedures. 

\begin{figure}[H]
    \centering
    \includegraphics[width=0.5\textwidth]{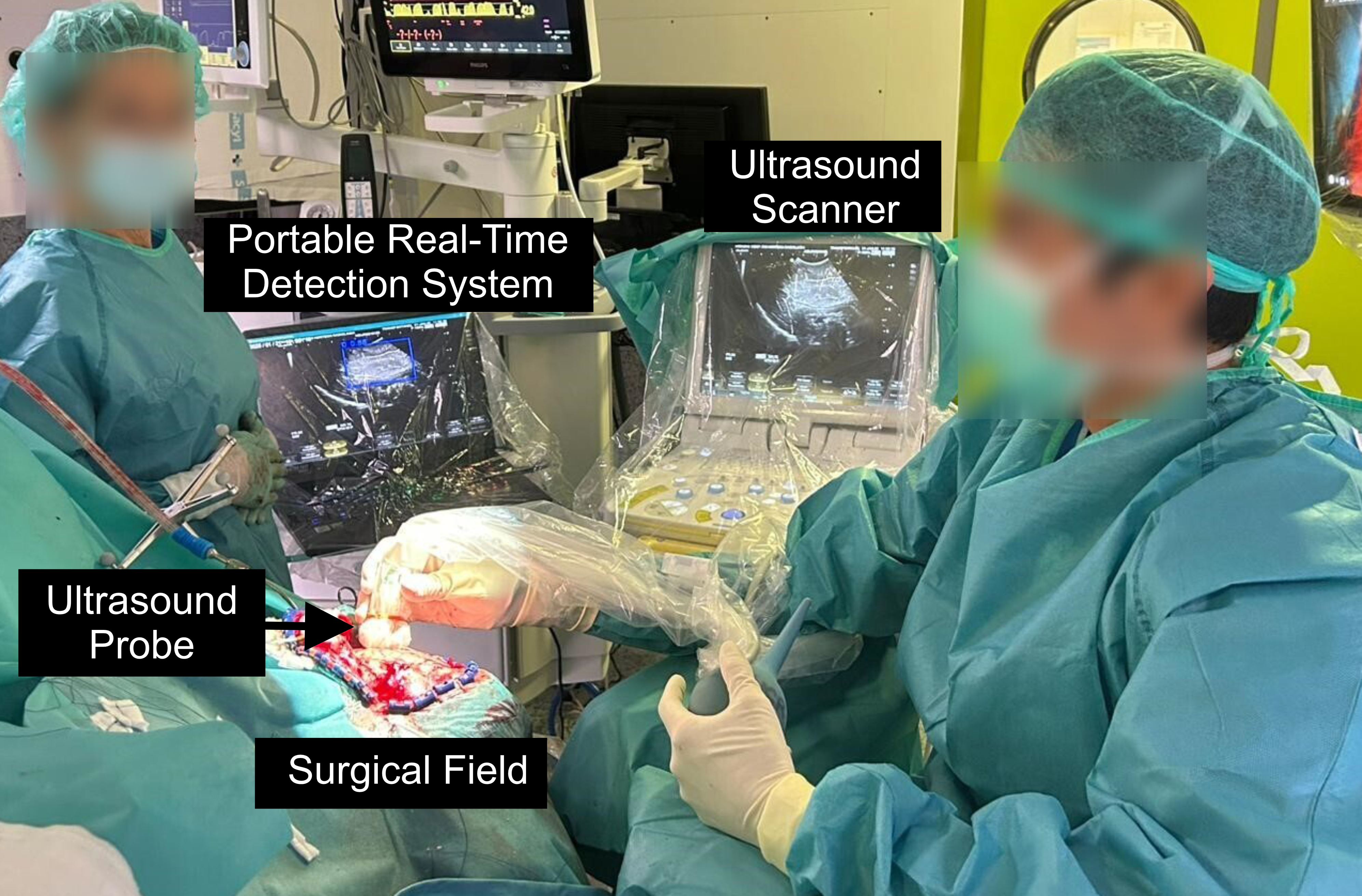}
    \caption{Operating room setup showing the integration of intraoperative ultrasound imaging, and the real-time object detection model using a portable prediction system.}
    \label{Figure_5}
\end{figure}

In all cases, the model was executed successfully, generating predictions that were considered correct by the observers. No latencies or delays were observed in the generation of the bounding boxes or in the visualization of the processed images. Although the original video signal from the ultrasound scanner contained additional elements, such as menu buttons and other visual components not present in the training images, the model effectively disregarded this nonrelevant information and generated accurate inferences. (Figure 6).

\begin{figure}[ht]
    \centering
    \includegraphics[width=0.5\textwidth]{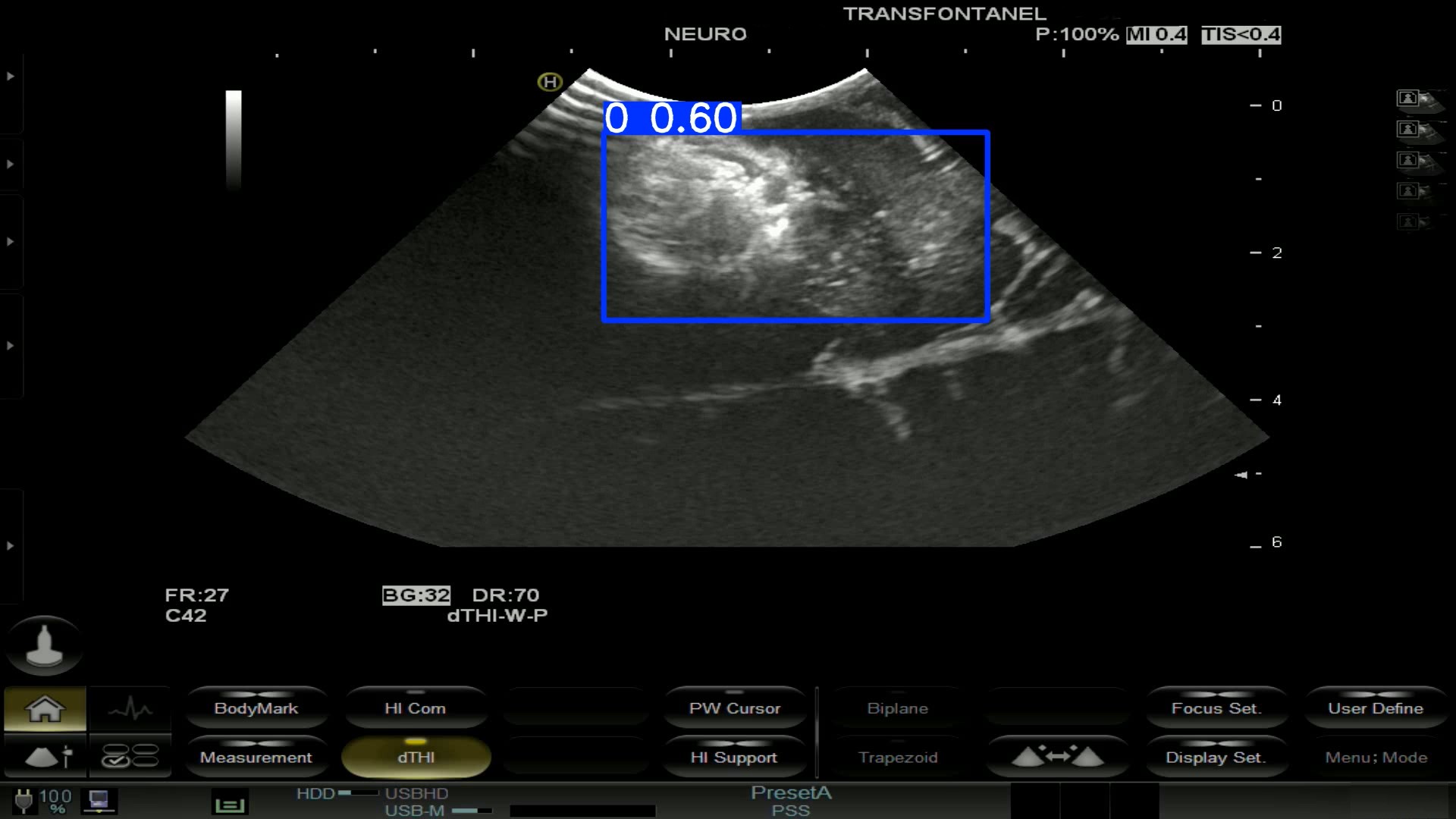}
    \caption{Example of a raw image captured from the video acquisition system in Case 8, showcasing a recurrent parasagittal meningioma. Note the elements surrounding the ultrasound field of view, including menus and buttons from the ultrasound scanner interface. In this case, the partially calcified tumor and surrounding edema are detected as a single region. Some elements in the image have been removed to protect sensitive information (e.g., patient identifiers)}
    \label{Figure_6}
\end{figure}

The surgical workflow was not interrupted at any time since the surgeons were able to work comfortably using both the computer screen dedicated to detection and the original screen of the ultrasound scanner. In the cases of low-grade gliomas, the model generated bounding boxes that correctly covered the entire tumor area, reflecting the echographic alteration characteristics of these tumors. In high-grade gliomas, the bounding box accurately delineates the tumor core, excluding regions with peritumoral alterations in echogenicity, such as edema or non-enhancing tissue.
Notably, despite the absence of images from other histological types in the training dataset, the model successfully detected tumors in patients diagnosed with meningiomas. However, in patients 9 and 10, the bounding box encompassed both the tumor and the perilesional edema, highlighting a limitation in distinguishing these regions in this specific tumor type (Figure 7).
In patients 2, 11, and 13, the model's ability to detect residual tumor was evaluated using images acquired during surgical resection. In patient 2, the model identified a small residual tumor in the insula (Figure 8A). For patient 13, despite using micro-Doppler mode, which was not included in the training dataset, the model successfully detected a residual tumor in the deep insular region adjacent to lenticulostriate arteries (Figure 8B). Although these types of images often contain surgery-associated artifacts, such as blood, hemostatic agents, and air, the model accurately identified the tumor areas, even in the presence of the surgical cavity. In patient 11, an ultrasound image acquisition was performed after completing the resection, during which the surgeon observed no visible macroscopic tumor. However, the model detected a small residual tumor region in one corner of the surgical cavity (Figure 8C). Upon further inspection, the surgeon confirmed that it was indeed an inadvertent tumor remnant. Another ultrasound image acquisition, performed after removing the residual tumor, confirmed the absence of residual tissue, as the surgeon verified it and the model did not detect any tumor (Figure 8D). These findings underscore the potential utility of the model in improving intraoperative detection and maximizing the extent of tumor resection.

\begin{figure}[H]
    \centering
    \includegraphics[width=0.4\textwidth]{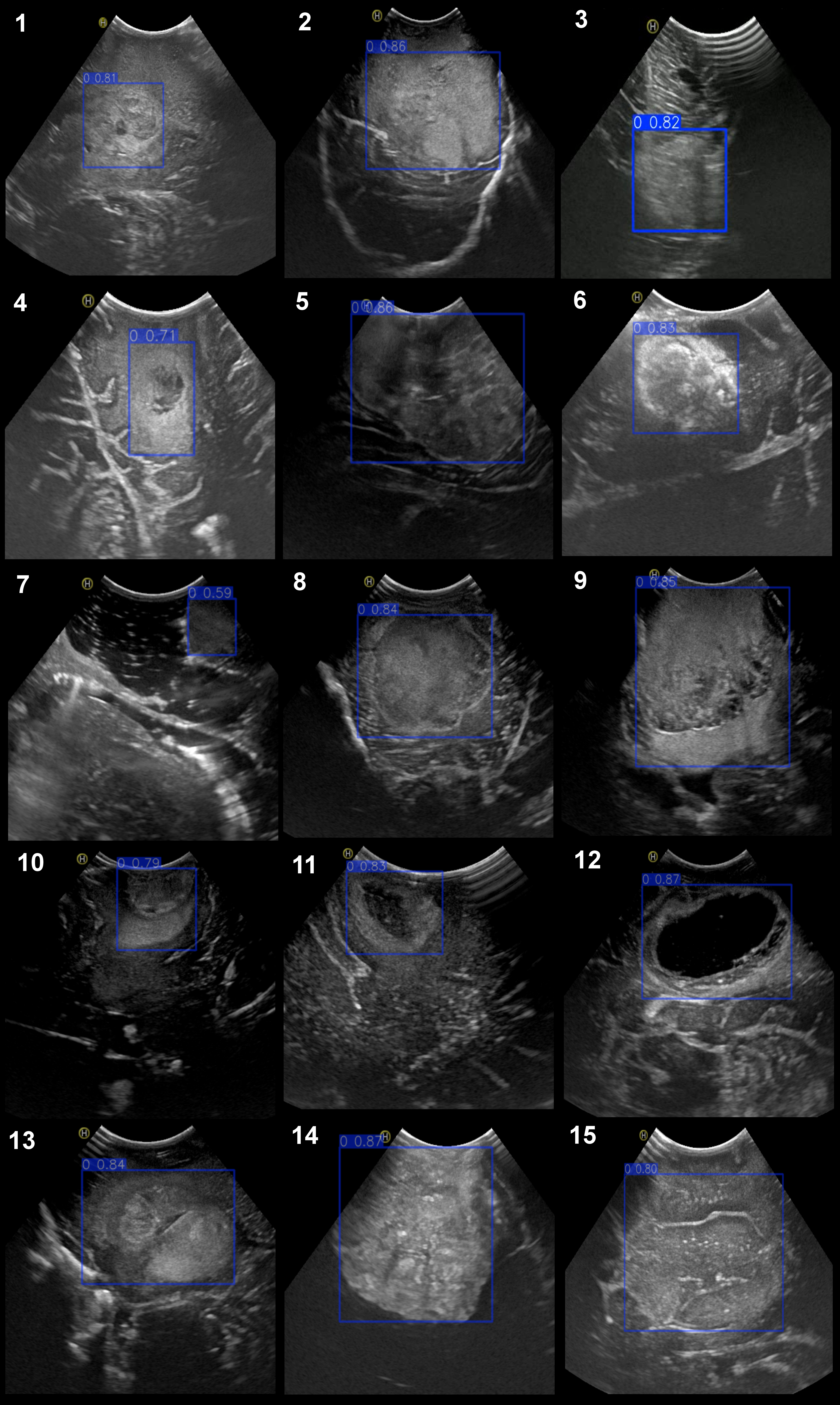}
    \caption{Real-time predictions of the YOLO11s detection model in prospective cases. The number in the top-left corner identifies the patient, corresponding to table 2. The blue bounding box represents the detection, and the accompanying value indicates the model's confidence score.}
    \label{Figure_7}
\end{figure}

\begin{figure}[ht]
    \centering
    \includegraphics[width=0.6\textwidth]{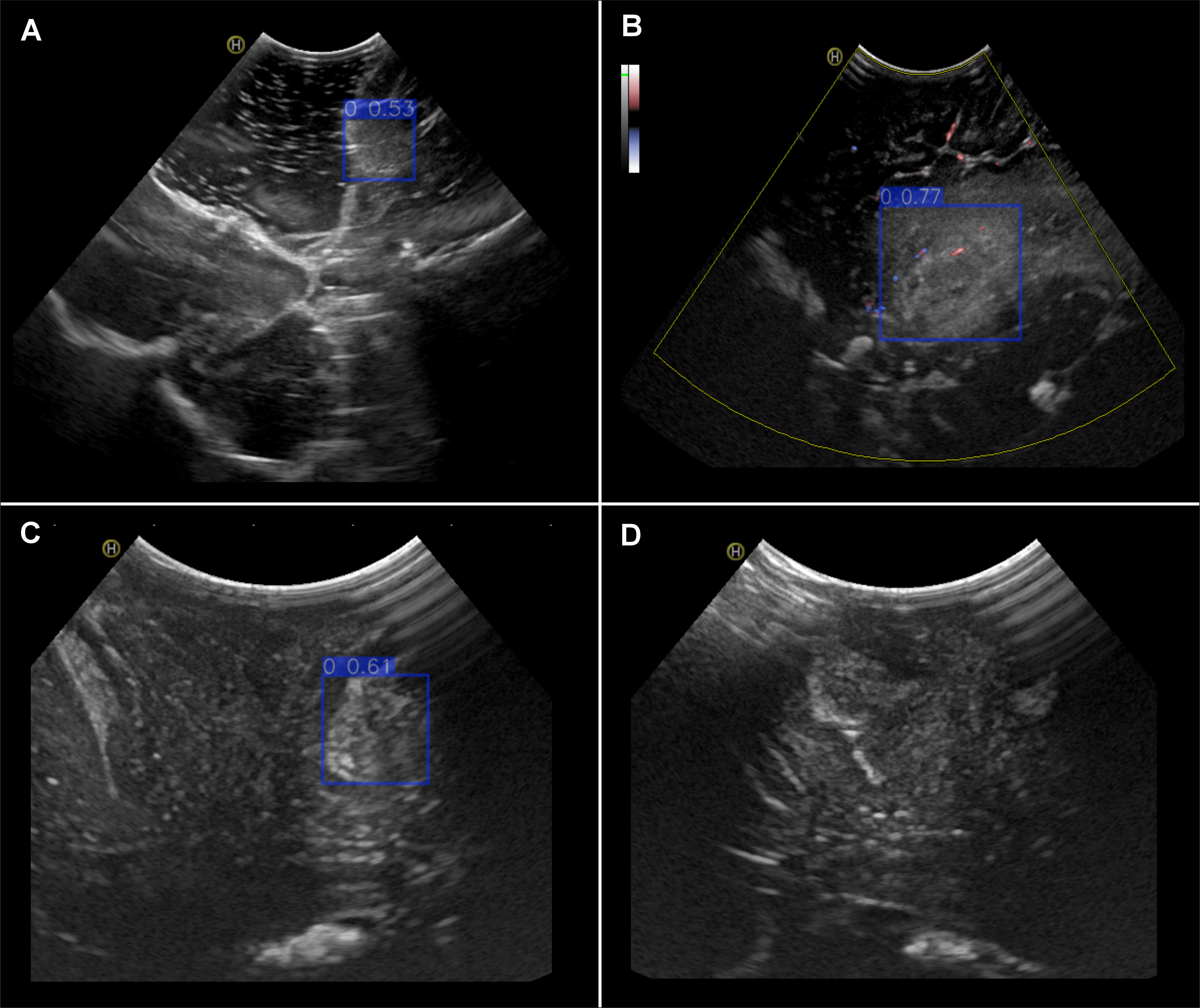}
    \caption{Examples of predictions in intraoperative images from prospective cases during surgical resection. (A) Patient 2: A small residual tumor is observed at the insular level. (B) Patient 13: Micro-Doppler mode image showing residual tumor at the deep insular level and the presence of lenticulostriate arteries. (C and D) Patient 11: In (C), a small residual tumor is observed, which is subsequently removed, as confirmed in (D), where no residual tumor is identified.}
    \label{Figure_8}
\end{figure}

\section{DISCUSSION}
\label{sec:discussion}
In this study, our primary objective was to train and evaluate a real-time brain tumor detection model, confirming its direct implementation in the operating room. To the best of our knowledge, this is the first study to apply an intraoperative object detection model in brain tumor surgery. The strengths and innovative nature of our approach are rooted in several key aspects.
First, our approach emphasizes object detection rather than segmentation, distinguishing it from previous studies. This choice aligns with specific clinical needs, where pixel-level precision is less critical, and rapid, straightforward detection is prioritized. Additionally, the dataset is composed primarily of high-quality native 2D images, providing a significant advantage over the 3D volumes or 2D-converted images from public datasets commonly utilized in earlier research.
The dataset, expanded to 11,570 images through data augmentation techniques, represents an unprecedented sample size in the field of brain tumors and intraoperative ultrasound, significantly enhancing the model's generalizability to real clinical scenarios. For the first time, an architecture such as YOLO11, optimized for both precision and computational efficiency and specifically tailored for real-time surgical applications, has been employed.
Both the detection and segmentation models developed in this study rely exclusively on 2D ultrasound images, removing the need for coregistration or fusion with preoperative magnetic resonance images, predefined segmentations, or manual intervention. This autonomous and streamlined approach simplifies clinical implementation and underscores the model's potential to transform intraoperative brain tumor management by providing precision and efficiency in the dynamic surgical environment.
In the detection task, which is the primary objective of this study, the results demonstrate that the different variants of YOLO11 achieve a balance between precision and computational efficiency, adapting to various clinical scenarios. For instance, YOLO11s achieves an mAP@50 of 0.95, an mAP@50-95 of 0.65, a latency of just 24.9 ms, and a speed of 34.16 FPS, making it highly suitable for real-time operating room environments. While more advanced models, such as YOLO11l (mAP@50-95 of 0.69) and YOLO11m (mAP@50-95 of 0.66), excel in detecting medium (mAP@50-95 between 0.70 and 0.75) and large (mAP@50-95 between 0.66 and 0.71) objects, lighter versions like YOLO11n (latency of 30.3 ms, mAP@50-95 of 0.68) offer advantages for systems with limited computational resources. These findings highlight the importance of selecting a model based on the specific requirements of the surgical environment and the available hardware capabilities.
There is no prior work in the literature on object detection in ultrasound images of brain tumors, which means there are no references available for a direct comparative analysis. However, as mentioned in Section 2, several authors have developed models for the segmentation task, providing a foundation for comparison. Our results demonstrate that the YOLO11 s model for the segmentation of brain tumors ioUS imaging achieves a Dice score of 0.88 and an IoU of 0.82, ranking among the highest values reported in the literature for similar tasks. Comparatively, Carton et al. \cite{Carton2020} achieved a median Dice coefficient of 0.74 when a 3D U-Net model was used to segment tumors in ioUS images, whereas Angel-Raya et al. \cite{AngelRaya2021} reported Dice coefficients between 0.85 and 0.86 when semiautomatic methods dependent on the registry with preoperative MR images were used. In more recent works, Faanes et al. \cite{Faanes2024} obtained an average Dice coefficient of 0.62 when MRI annotations were transferred to ioUS, confirming the limitations of registry-based methods, especially for small tumors. In contrast, our model demonstrates robust and consistent performance for different tumor sizes, with a latency of 24.4 ms and a speed of 20.79 FPS, which are critical characteristics for its implementation in real time in the operating room.
On the other hand, our group has recently published a model based on brain tumor segmentation using part of the data from the BraTioUS and ReMIND datasets \cite{Cepeda2025}. This model reached a median DSC of 0.93 in an independent external cohort and 0.65 in the RESECT-SEG dataset, indicating good generalizability in diverse clinical settings. Compared with these results, the YOLO11 s model stands out for its balance between precision, computational efficiency and speed, presenting itself as an innovative and practical solution for automated real-time segmentation of brain tumors in ioUS images. These findings underscore the potential of our approach to overcome the challenges inherent in ioUS imaging and improve intraoperative precision in neurosurgical procedures.

\subsection{Limitations}
The main limitation of our work lies in the potential bias introduced by the manual segmentation used as a reference for the training of the model. Although this was performed by two expert observers in neuro-oncology and intraoperative ultrasound, some interobserver variability is unavoidable, a phenomenon widely documented in the literature, as described by Weld et al \cite{weld2024challenges}. This bias could influence the precision of the model, especially in cases where the boundary delimitation is more ambiguous.
In particular, low-grade gliomas represent a significant challenge because of their subtle ultrasound characteristics and the lack of clear contrast between the tumor tissue and the surrounding healthy tissue. These intrinsic challenges contribute to segmentation uncertainty and reduce annotation consistency, leading to inherent variability that is unavoidable in such cases. Overcoming these limitations in future studies will require refining annotation strategies and developing more robust models capable of managing the variability associated with manual segmentation.

\subsection{Future directions}
Our work, although innovative, is susceptible to improvements and extensions. Despite the remarkable efficiency demonstrated by YOLO11 in this and other areas, it is important to explore other architectures or modifications that may offer superior performance or better adaptation to the specific characteristics of intraoperative ultrasound images.
A possible line of improvement would be the inclusion of annotations not only of the tumor region but also of the peritumoral region (edema) and other cerebral anatomical structures, such as the ventricles and the falx cerebri. This could further improve the interpretability of the images and provide a richer anatomical context, facilitating the integration of the model in surgical practice.
Another promising direction would be to include images acquired during tumor resection in training, placing special emphasis on small tumor remnants. This could significantly increase the usefulness of the model as an auxiliary tool to achieve better extent of resection rates, especially in cases where the residual tumor may be unnoticed by the human eye.
Finally, implementing and prospectively validating the model in other centers, involving neurosurgical teams from different institutions and using ultrasound equipment from various manufacturers, is essential to assess its generalizability and robustness across diverse clinical settings. These improvements and validations could expand the scope and applicability of our work, consolidating it as a versatile and effective tool in intraoperative ultrasound-assisted surgery.

\section{CONCLUSION}
\label{sec:conclusion}
Intraoperative ultrasound plays a crucial role as an imaging modality in brain tumor surgery, and its optimization can be significantly improved by the integration of computer vision techniques. These tools have the potential to contribute to better image interpretation and precise tumor localization during surgical procedures.
In this study, we trained a brain tumor detection model using the YOLO11 architecture. The results obtained demonstrate outstanding performance in terms of accuracy and computational efficiency, which are fundamental aspects for its application in real time. The successful implementation of the model in the operating room confirms its viability as a practical and robust tool, with enormous potential to optimize surgery for this type of neoplasm. This work lays the foundation for the development of advanced surgical support systems that can improve clinical outcomes and facilitate the work of neurosurgical teams.

\section*{FUNDING}
This work was funded by the grant GRS 2313/A/21, titled “Prediction of overall survival in glioblastomas using radiomic features from intraoperative ultrasound: A proposal for the creation of an international database of brain tumor ultrasound images,” awarded by the Gerencia Regional de Salud de Castilla y León.

\section*{CONFLICTS OF INTEREST}
B.V.N. serves as a consultant for BK Ultrasound, Brainlab, React Neuro, and Robeuate.

\printbibliography

\end{document}